\documentclass[onecollarge,natbib]{svjour2}
\bibpunct{[}{]}{,}{n}{}{,} 
\smartqed  
\usepackage{graphicx}
\def\beq{\begin{equation}}
\def\eeq{\end{equation}}
\def\bea{\begin{eqnarray}}
\def\eea{\end{eqnarray}}
\def\beqa{\begin{equation}\begin{array}{l}}
\def\eeqa{\end{array}\end{equation}}
\def\eqlab#1{\label{eq:#1}}

\def\Eqref#1{Eq.~(\ref{eq:#1})}

\def\Figref#1{Fig.~\ref{fig:#1}}

\def\barr{\left(\begin{array}{c}}
\def\earr{\end{array}\right)}
\def\bmat{\left(\begin{array}{cc}}
\def\emat{\end{array}\right)}

\def\ga{\gamma}

\def\la{\lambda}

\def\mathscr{\mathcal}

\def\3d{3-D}

\journalname{Few-Body Systems (APFB2011)}
\begin{document}

\title{\boldmath
Causality in the relativistic bound-state problem
}
\subtitle{In memory of John A.\ Tjon}

\author{Vladimir Pascalutsa }


\institute{V.~Pascalutsa \at
              Institut f\"ur Kernphysik \\
              University of Mainz\\
              55099 Mainz, Germany\\
              \email{vladipas@kph.uni-mainz.de}           
}

\date{Received: date / Accepted: date}

\maketitle

\begin{abstract}
Although the exact Bethe-Salpeter equation is certainly the appropriate field-theoretic framework
to describe the non-perturbative problem of scattering and bound states, the inevitable truncations
introduce inconsistencies such as loss of symmetries or incorrect one-body limit. 
I conjecture that these problem can be overcome if the truncation preserves the field-redefinition invariance
of the exact equation. A sum rule for light-by-light
scattering can provide a testing ground of this conjecture.

\end{abstract}

\section{Introduction}
\label{rem}

John Tjon has had about 30 PhD students over
the course of his remarkable life in physics.
I am privileged to be one of them,  one of the very last in fact.
In these proceedings I shall go through 
a piece of physics I was learning from John.
It concerns the bound-state problem in relativistic theory, or, 
more specifically, the quest for a
relativistic analog of the Lippmann-Schwinger equation
which would reduce to the appropriate one-body equation
(Klein-Gordon or Dirac, depending on the spin) in the limit when
one of the particles is infinitely heavy, viz., one-body limit. 
This is quite an old and rich subject and I would
never be able to present it fairly, but the bottom line is that
the only fully consistent equation is the exact Bethe-Salpeter equation (BSE) \cite{BeS51}, 
derived from quantum field theory \cite{ItZ88}. 
The problem with the BSE is that in practice it can not be solved exactly because the kernel
consists of infinitely many two-particle irreducible (2PI) graphs.  One needs to truncate, and that is very difficult to do in a systematics fashion. 
For example, keeping only the trees (one-particle exchanges) 
in the kernel leads to the ladder BSE which does not
have the correct one-body limit. 

An improvement of the kernel, e.g., the inclusion of
the 2PI one-loop graphs, presumably improves the situation. But in the earlier days of nuclear physics it was not clear how to include such loops, 
which in the case of the nucleon-nucleon ($NN$)  interaction, for example, 
would be entering with higher orders of the large $\pi NN$ coupling.
It is only with the advent of the effective-field theories (EFTs) \cite{Wei79}, that we have
learned how systematically expand the $NN$ kernel in powers of small
energy scales \cite{Weinberg:1991um,vanKolck:1994yi,Epelbaum:1998ka,Epelbaum:2010nr}.
To date, however, just a handful of the EFT calculations in few-nucleon
systems include relativistic effects, let alone attempt to solve
the full BSE. It is usually argued that relativistic effects
are negligible because the nucleon 
mass is much greater than the relative energy
of nucleons in nuclei. I am not very happy with this argument, because by 
a similar argument the heavy-baryon expansion (HBChPT) in the single-nucleon sector would always converge well, and we know it is not, e.g.~\cite{Becher:1999he,Pascalutsa:2004ga}. 
In fact, calculations in the single-nucleon sector show that by the time the $\Delta(1232)$-resonance
excitation becomes important relativity kicks in \cite{Pascalutsa:2006up}.  
A dramatic example is provided by the magnetic polarizability of the nucleon, where the $\Delta$-excitation and relativistic effects largely cancel each other, leaving us with the small value for this quantity
\cite{Lensky:2009uv}. 
As the $\Delta(1232)$-isobar has  recently been
included in the effective $NN$ potential and shown to play an important role \cite{Krebs:2011zz}, further
relativistic EFT studies will hopefully be done in the near future.

\section{What Gross equation and infrared regularization of BChPT have in common}
\begin{figure}[h]
\centering
  \includegraphics[width=0.9\textwidth]{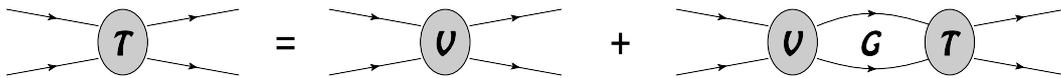}
\caption{Graphical representation of a two-body scattering equation.}
\label{fig:1}       %
\end{figure}

Going back to the pre-EFT era, when the loops were out of question,
it became customary to improve the kernel by an additional approximation, however paradoxically that might sound. The additional approximation is 
called quasi-potential reduction of the Bethe-Salpeter
equation (BSE). If we write BSE, for the
scattering amplitude in momentum space, as (see \Figref{1}):
\beq
\eqlab{BSE}
T(p',p) = V(p',p) + i\int \frac{d^4 k }{(2\pi)^4} V(p', k)\,  G(k)\,  T(k,p), 
\eeq 
where $VG$ is the kernel, then the 3-dimensional reduction is done by
manipulating the dependence of $VG$, and subsequently $T$, on $ k_0$,
the energy component of the integration four-momentum. It is assumed 
that $k_0$ is not an independent variable, but is fixed
some way in terms of the external momenta and possibly $\vec k$.
Any choice of fixing $k_0$ is okay as long as the unitarity and Lorentz
invariance are not violated.
This procedure  lifts  the integration over $k_0$, leaving us 
with a much simpler 3-dimensional equation. Sometimes a quasi-potential 
equation is just introduced  or postulated, and has no evident 
connection to BSE, but still, they all are 3-dimensional and we refer to them 
as `relativistic 3D equations'.

In 1969 Franz Gross described a relativistic 
3D equation \cite{Grs69}, bearing now his name, which has the
correct one-body limit for one-boson-exchange potential. It thus provides a
consistent formulation of the relativistic Yukawa problem, but with some caveats. 
The Gross equation is
obtained in a 3D reduction of the ladder BSE  by putting the heavier particle on the mass shell.
This means that in doing the $k_0$ integration in \Eqref{BSE} by counting
the poles in the complex $k_0$ plane, one takes the contribution of one and only
one pole: the positive-energy pole of the heavy-particle propagator, contained
in the two-particle propagator $G$. The contribution of all the other poles
in either $V$, $G$, or $T$ is discarded. This seems as a harsh approximation,
but it does the trick with the one-body limit, cf.\ Chapter 12 in \cite{Grs94} for more details. 

The first caveat is that the one-body limit is recovered for one-neutral-boson-exchange potentials only \cite{Pascalutsa:1999pv}, which makes it of limited help in practice but is not a conceptual problem. 
The caveat that concerned John Tjon, during my work with him, was
the apparent lack of charge conjugation symmetry \cite{PT97}.  
It is interesting, although perhaps obvious to some,
that while the equation is Lorentz-invariant, the loss
of charge-conjugation symmetry leads to violation of (micro-) causality.
The relativistic covariance is thus compromised at the quantum level ---
virtual states may propagate outside the light cone.
It is clear that the resulting unphysical effects must diminish in the
strict one-body limit, but it is hard to assess their size away from the limit, where
the equation is actually applied. 

Years later I encountered a similar situation in the infrared regularization (IR) of 
baryon chiral perturbation theory (BChPT), 
introduced by Becher and Leutwyler in 1999 \cite{Becher:1999he}. The IR is a scheme
to calculate chiral loops with nucleons such that no positive powers of the nucleon mass $M_N$
occur --- the positive powers were thought to violate the power counting of BChPT.\footnote{When
being a postdoc at Flinders University in Adelaide between 1999 and 2001, 
I learned from a graduate student at our department (Jambul Gegelia) that the positive powers
of $M_N$ do not break power counting because their effect is always absorbed by low-energy
constants \cite{Gegelia:1999qt}. But note that \cite{Gegelia:1999qt} first appeared in 1999
and was only published in 2003. It took some time before this work was considered
as a legitimate solution to the power-counting problem.}
The original formulation of IR is done in terms of Feynman parametrization of loop integrals,
but when checking the Ward-Takahashi identities in this scheme, I found that all the IR does is to
remove the negative-energy pole of nucleon propagators. Namely, if
we consider a one-loop  graph with one pion propagator,
$
S_\pi(k) = (k^2 -m_\pi^2)^{-1},
$ and any number of nucleon propagators, 
$
S_N(p) = (\ga\cdot p - M_N)^{-1} ,
$
 then as the result of the IR procedure every
nucleon propagator is replaced as follows:
\beq
S_N(p) \stackrel{\mathrm{IR}}{\to} S_N(p) \Big(1+\frac{1}{S_\pi(k) (p^2-M_N^2)} \Big)
= \frac{\ga\cdot p + M_N}{p^2-k^2 - M_N^2+m_\pi^2 }
\eeq
Since both $p$ and $k$ depend linearly on the loop momentum, the denominator of the modified
nucleon propagator is linear too, the quadratic term cancels. Hence this nucleon propagator has
only one pole, the positive-energy one. 
This procedure has been generalized by Tim Ledwig to any number of pion propagators \cite{Ledwig:2011cx}. Graphs with
no pion propagators vanish in IR, since all the poles lie in the same half-plane of the complex energy plane.
 
 So, once again, deleting the negative-energy pole violates charge-conjugation symmetry, hence causality.
 It is manifested in the unphysical cuts  whose appearance  was already noticed in the
 original work \cite{Becher:1999he}. It is argued nonetheless that the cuts lie far away from the
 domain of interest of BChPT, and hence their effect should be negligible. Unfortunately this argument
 is not always corroborated in actual calculations, e.g.~\cite{Pascalutsa:2004ga,Holstein:2005db,Geng:2008mf,Alarcon:2011vm}.

To summarize thus far, the Gross equation as well as IR-BChPT neglect the negative-energy nucleon
pole in the loop calculations. This seems as a very safe approximation at low energies, since
the nucleon is rather heavy and it costs a lot of energy to produce an anti-nucleon. 
However, in loops we integrate over all energies, and
the absence of virtual anti-nucleons induces the unphysical (acausal) high-energy contributions.
The net effect of these contributions on low-energy physics
is expected to be small, but cannot be assessed a priory. In EFT the high-energy physics probed
in the loops is compensated by low-energy constants, but in order to use this argument in the IR case
one needs to use an effective Lagrangian without the charge-conjugation symmetry.

\section{Sum rules for light-by-light scattering as a test of causality}
\label{sec:1}

In a given calculation it is usually not difficult to check gauge invariance or unitarity. A test of causality
is less obvious,  but a test at the final stage of  a calculation can be provided by
sum rules such as GDH.  The sum rules involve cross sections, the observables which
should be computable in any physical theory, even the string theory. The GDH sum rule itself involves
other quantities, such as the anomalous magnetic moment, which might obscure the test.
However, there is at least one exact sum rule which involves cross sections only:
\beq
\eqlab{LbL}
\int_0^\infty\! \frac{d s}{s} \Big[ \sigma_2 (s) - \sigma_0(s)\Big] = 0,
\eeq
where $\sigma_\la $ are polarized total cross sections for $\ga\ga$ fusion into anything; $\sigma_0$
is for circularly polarized photons with the same helicity, and $\sigma_2$ with opposite. The integration
is over the total invariant energy, the Mandelstam variable $s$.
This sum rule is derived from general properties of unitarity,
analyticity, crossing, and gauge invariance of the light-by-light scattering amplitude \cite{Pascalutsa:2010sj}, 
and as such it provides a test ground of these principles.
\begin{figure}[b]
\centering
  \includegraphics[width=0.9\textwidth]{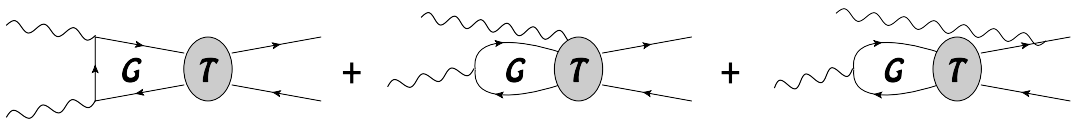}
\caption{Graphical representation of the $\ga\ga \to N\bar N$ amplitude. Crossed graphs are omitted.}
\label{fig:2}       %
\end{figure}

 To perform this test for a relativistic scattering equation one should first be able to construct
 the particle-antiparticle scattering amplitude, e.g. $N\bar N$ if we talk about the nucleons.
 Based on that, one should obtain the gauge-invariant $\ga\ga \to N\bar N$ amplitude (see \Figref{2}),  
 calculate the corresponding polarized cross sections, and integrate them as in \Eqref{LbL}
 to find zero, or not. In the latter case, one can compare the result with the size of the cross sections.
 If the integral is much smaller than cross sections themselves, the violation is small.

\section{Field-redefinition invariance}

The two-body scattering  equations have issues bigger than causality. 
Fundamental local symmetries, such as the e.m.\ 
gauge invariance, are not easy to maintain either, see e.g., \cite{GrR87,GrS93}. 
Even the non-relativistic EFT calculations face this problem when the photon comes in,
e.g., in the description of electron-deuteron scattering. And even without the photons,
in plain Lippmann-Schwinger description of $NN$ (with non-perturbative pions), because the chiral symmetry is obscured by iterations and it is not clear how to renormalize \cite{Epelbaum:2010nr}.   

It seems the source of all these issues lies in field-redefinition invariance (FRI). Or more precisely, in 
the fact that the equations are not invariant under field redefinitions. 

In quantum field theory the FRI is a simple consequence of independence 
of the partition function on the choice of integration variables. In a two-body equations,
this is much less trivial, since already the separation of graphs into reducible and irreducible
is not invariant. Of course when all graphs are present the answer is invariant, but, if truncations
are made as they must,  it is generally not. 

To see an example, consider the leading order  $NN$ interaction in the chiral effective theory of
the nuclear force (in Weinberg's counting \cite{Weinberg:1991um}). 
The leading-order potential consists of a four-nucleon contact interaction and 
a pion exchange with the $\pi NN$ vertices bearing the pseudo-vector coupling. 
After making a redefinition of
the nucleon field as \cite{Lensky:2009uv}:
\beq
N(x)  \to \exp{\Big( i \frac{g_A }{ 2 f_\pi} \tau_a \pi^a (x)\,  \ga_5 \Big)} \, N(x),
\eeq
where $\pi(x)$ is the pion field, we can see that the only change in the potential is that the pseudo-vector
$\pi NN$ coupling is replaced by the pseudo-scalar one. The on-shell potential has not changed,
but the off-shell one has, and hence the solution changes. The change is
quite dramatic, e.g., the pion exchange contribution is finite and one does not need a regulator
(i.e., the finite cutoff) which otherwise is always present in such calculation \cite{Epelbaum:2010nr}.
The bottom line is that there is a strong dependence of the result on the choice of the nucleon field,
a feature which is not desirable given the fact that the FRI is assumed in constructing the chiral Lagrangian.

Finding a truncation of BSE which maintains the field-redefinition invariance, will solve
many, if not all, of the aforementioned problems. Whether it can be done in principle is
another question.

\begin{acknowledgements}
It is a pleasure to thank Franz Gross and Evgeny Epelbaum for delightful discussions
during the conference. 
\end{acknowledgements}



\end{document}